\documentclass[mathleft,fleqn,%
]{an}
\usepackage{graphicx}
\usepackage[varg]{txfonts}
\begin{document}


\Pagespan{1}{}
\Yearpublication{2014}%
\Yearsubmission{2014}%
\Month{0}%
\Volume{999}%
\Issue{0}%
\DOI{asna.201400000}%

\title{
 Chemical spots on the surface of the strongly magnetic Herbig Ae star HD\,101412\thanks{Based on data
   obtained from the ESO Science Archive Facility under request MSCHOELLER~101895
   (ESO programme Nos.~081.C-0410(A), 085.C-0137(A), and 187.D-0917(D)).}}

\titlerunning{Chemical spots on the surface of HD\,101412}

\author{S.\,P.  J\"arvinen\inst{1}\fnmsep\thanks{E-mail:
        {sjarvinen@aip.de}}
\and S. Hubrig\inst{1}
\and M. Sch\"oller\inst{2}
\and I. Ilyin\inst{1}
\and T.\,A. Carroll\inst{1}
\and H. Korhonen\inst{3,4}
}

\authorrunning{J\"arvinen et al.}

\institute{
Leibniz-Institut f\"ur Astrophysik Potsdam (AIP), 
An der Sternwarte~16, 14482~Potsdam, Germany
\and
European Southern Observatory, 
Karl-Schwarzschild-Str.~2, 85748~Garching, Germany
\and
Finnish Centre for Astronomy with ESO (FINCA), University of Turku, 
V\"ais\"al\"antie~20, 21500 Piikki\"o, Finland
\and
Dark Cosmology Centre, Niels Bohr Institute, 
Copenhagen University, Juliane Maries Vej 30, 2100, Copenhagen {\O}, Denmark
}

\received{2015}
\accepted{2016}
\publonline{XXXX}


\abstract{
Due to the knowledge of the rotation period and the presence of a
rather strong surface magnetic field, the sharp-lined young Herbig Ae
star HD~101412 with a rotation period of 42\,d has become one of the
most well-studied targets among the Herbig Ae stars. High-resolution
HARPS polarimetric spectra of HD\,101412 were recently obtained on
seven different epochs. Our study of the spectral variability over the
part of the rotation cycle covered by HARPS observations reveals that 
the line profiles of the elements Mg, Si, Ca, Ti, Cr, Mn, Fe, and Sr
are clearly variable while He exhibits variability that is opposite to
the behaviour of the other elements studied. Since classical Ap stars 
usually show a relationship between the magnetic field geometry and
the distribution of element spots, we used in our magnetic field
measurements different line samples belonging to the three elements
with the most numerous spectral lines, Ti, Cr, and Fe. Over the time
interval covered by the available spectra, the longitudinal magnetic
field changes sign from negative to positive polarity. The
distribution of field values obtained using Ti, Cr, and Fe lines is,
however, completely different compared to the magnetic field values 
determined in previous low-resolution FORS\,2 measurements, where 
hydrogen Balmer lines are the main contributors to the magnetic field 
measurements, indicating the presence of concentration of the studied 
iron-peak elements in the region of the magnetic equator. Further, we 
discuss the potential role of contamination by the surrounding warm 
circumstellar matter in the appearance of Zeeman features obtained
using Ti lines. 
}

\maketitle

\section{Introduction}

Among the Herbig stars with detected magnetic fields, HD\,101412 possesses 
the strongest magnetic field up to 3.5\,kG 
(Hubrig et al.\ \cite{Hubrig2010}), 
followed by V380\,Ori with a dipole field strength of about 2.1\,kG 
(Alecian et al.\ \cite{alecian2009}). 
The presence of such a strong magnetic field on the surface of HD\,101412 
makes it a prime candidate for high-resolution spectropolarimetric studies of 
the impact of the magnetic field on the physical processes occurring during 
stellar formation. 
Hubrig et al.\ (\cite{Hubrig2009}) 
tested a number of atmospheric models in the range: 
$T_{\rm eff}=8000-11\,000$\,K and $\log g=4.0-4.05$ and obtained as a best 
fit $T_{\rm eff}=10\,000\,K$, $\log g=4.2-4.3$, and a $v \sin i$ value of 
about 5\,km\,s$^{-1}$. This result is in good agreement with the study of 
Guimar\~aes et al.\ (\cite{Guimaraes2006}), who published 
$T_{\rm eff}=10\,000\pm1000$\,K, $\log g=4.1\pm0.4$, and 
$v\,\sin\,i=7\pm1$\,km\,s$^{-1}$. A good fit was also obtained using the 
model $T_{\rm eff}=9000$\,K and $\log g=4.0$, but the synthetic spectrum for 
these parameters showed many narrow lines that did not appear in the observed 
UVES spectra (Hubrig et al.\ \cite{Hubrig2009}). In their work, the authors 
mentioned that the spectrum of 
HD\,101412 is heavily contaminated by weak circumstellar (CS) lines. The 
fundamental parameters, $T_{\rm eff}=8\,300$\,K and $\log\,g=3.8$, and even 
lower projected rotation velocity, $v\,\sin\,i=3\pm2$\,km\,s$^{-1}$, were 
determined by Cowley et al.\ (\cite{cowley2010}) using high-resolution HARPS spectra. 
According to their results, the photosphere is depleted in most refractory 
elements, while volatiles are normal or, in the case of nitrogen, overabundant 
with respect to the Sun. The authors suggested that the detected anomalous 
saturation of strong lines arises from heating of the upper atmospheric layers 
by infalling material from a disk. The overall abundance pattern may be 
related to those found for $\lambda$~Boo stars.

As of today, only about 20 late Herbig~Be and Herbig~Ae stars have been 
reported to possess large-scale organized magnetic fields 
(e.g.\ Hubrig et al.\ \cite{Hubrig2009}, \cite{Hubrig2015}; Alecian et al.\ \cite{alecian2013}), 
using low- and/or high-resolution spectropolarimetric observations, and only 
for about half a dozen of this type of stars the detection of the magnetic 
field was achieved using both low- and high-resolution spectropolarimetric 
observations. Moreover, only for the two Herbig~Ae stars HD\,101412 and 
V380\,Ori 
(Hubrig et al.\ \cite{Hubrig2011a}; Alecian et al.\ \cite{alecian2009}), 
the magnetic field geometry has been constrained in previous studies. 

Previously published mean longitudinal magnetic field measurements of the 
Herbig Ae star HD\,101412 were based on low-resolution polarimetric spectra 
obtained with FORS\,1/2 (FOcal Reducer low dispersion Spectrograph) mounted on 
the 8-m Antu telescope of the VLT 
(e.g.\ Wade et al.\ \cite{Wade2005}, \cite{Wade2007}; 
Hubrig et al.\ \cite{Hubrig2009}, \cite{Hubrig2010}, \cite{Hubrig2011a}). 
In contrast to the results obtained by 
Wade et al.\ (\cite{Wade2007}) 
who measured a positive magnetic field of the order of 500\,G using hydrogen 
lines and simultaneously on the same spectra a negative 
magnetic field of the same order using metal lines, the 
studies of Hubrig et al.\ showed rather consistent results between the measurements 
obtained using the full spectrum including metal lines and those using hydrogen lines.

Combining photometric observations and measurements of the longitudinal magnetic 
field based on FORS\,2 spectra obtained on 13 different epochs distributed 
over 62 days, 
Hubrig et al.\ (\cite{Hubrig2011a}) 
determined for HD\,101412 a rotation period of 42.076\,d. The authors report 
that HD\,101412 exhibits a single-wave variation in the longitudinal magnetic 
field during the stellar rotation cycle, which is usually considered as 
evidence for a dominant dipolar contribution to the magnetic field topology. 
Furthermore, high-resolution, high signal-to-noise UVES spectra and a 
few lower quality HARPS spectra revealed the presence of a few resolved 
magnetically split lines and a variation of the mean magnetic field modulus 
(Hubrig et al.\ \cite{Hubrig2010}). 
The Zeeman doublet \ion{Fe}{ii} at $\lambda$~6149.258 appeared resolved in the 
acquired spectra with the measured mean magnetic field modulus varying from 
2.5 to 3.5\,kG, while the mean quadratic magnetic field was found to vary in 
the range from 3.5 to 4.8\,kG. The mean quadratic magnetic field is determined 
from the study of the second-order moments of the line profiles recorded in 
unpolarised light (that is, in the Stokes parameter $I$). Such an analysis is 
usually based on the consideration of samples of reasonably unblended lines 
in spectra (e.g.\ Mathys \& Hubrig \cite{MathysHubrig2006}). 
Noteworthy, HD\,101412 is presently the only Herbig Ae star for which the 
rotational Doppler effect was found to be small in comparison to the magnetic 
splitting of several spectral lines observed in unpolarised light.

High-resolution spectropolarimetric observations of HD\,101412 on
seven different epochs obtained using the HARPS (High Accuracy 
Radial velocity Planet Searcher) spectropolarimeter are publicly available 
in the archive (Prg.~187.D-0917) of the European Southern Observatory (ESO). 
In contrast to low-resolution spectropolarimetry, the high-resolution 
polarimetric spectra allow us to study in detail surface abundance 
inhomogeneities and, in particular, how different elements with different 
abundance distributions across the stellar surface sample the magnetic field. 
The same spectra were used by 
J\"arvinen et al.\ (\cite{jarvinen}),
where magnetic field measurements using iron lines were presented 
along with the magnetic field measurements of two other sharp-lined 
Herbig Ae stars, HD\,104237 and HD\,190073. The emphasis in that study
of HD\,101412 was however put on the investigation of the impact of
the circumstellar matter on the observed line profiles.

In Sect.~\ref{sect:obs}, we report on the observations and 
data reduction, in Sect.~\ref{sect:variab} we present our search for spectral 
variability using lines belonging to different elements, and in 
Sect.~\ref{sect:mf_meas} we describe the methods and results of our magnetic 
field measurements using samples of lines belonging to different elements. 
Finally, in Sect.~\ref{sect:disc} we discuss the significance of the obtained
results for future directions in the studies of Herbig Ae stars.


\section{ Observations and data reduction }\label{sect:obs}

\begin{table}
\begin{center}
\caption{Logbook of the HARPS spectropolarimetric observations of the 
Herbig~Ae star HD\,101412.
}
\label{tab:log_meas}
\begin{tabular}{ccrc}
\hline\hline
HJD & Phase\footnote{} & SNR & Exp.\ time\\ 
2450000+ &  &  & [s]\\
\hline
6337.801 & 0.131 & 101 & $4\times700$\\
6337.835 & 0.132 & 103 & $4\times700$\\
6338.832 & 0.156 & 104 & $4\times700$\\
6339.709 & 0.176 &  92 & $4\times600$\\
6341.732 & 0.225 &  76 & $4\times600$\\
6342.634 & 0.246 &  58 & $4\times600$\\
6344.782 & 0.297 &  60 & $4\times600$\\
\hline
\end{tabular}
\end{center}
$^{1}$ The phases are based on the rotation period of 42.076\,d determined by 
Hubrig et al.\ (\cite{Hubrig2011a}).
\end{table}

Seven spectropolarimetric observations were obtained with the HARPS 
polarimeter (HARPSpol; 
Snik et al.\ \cite{snik2008}) 
attached to ESO's 3.6\,m telescope (La~Silla, Chile) on the nights from 2013 
February 14 to 21. Among them, two individual observations were obtained 
during the first night on February 14. Although these two observations are 
obtained at approximately the same rotation phase, we consider them in our 
magnetic field study individually to learn about the limitations of our 
measurement methods. The seven HARPS archival spectra cover rotation phases 
from 0.131 to 0.297. Each observation was split into four subexposures with an 
exposure time of 10--12\,min, obtained with different orientations of the 
quarter-wave retarder plate relative to the beam splitter of the circular
polarimeter. The achieved signal-to-noise ratio (SNR) in the final Stokes~$I$ 
spectra, summed over the four subexposures, is rather low, accounting for a SNR 
between 58 and 104. In Table~\ref{tab:log_meas}, the dates of observations are 
presented in the first column, followed by the rotation phase and the 
achieved signal-to-noise ratio. The exposure times are given in the last 
column. The spectra have a resolving power of about $R = 115\,000$ and cover
the spectral range  3780--6910\,\AA, with a small gap between 5259 and 
5337\,\AA. The reduction and calibration of the archive spectra was performed 
using the HARPS data reduction software available at the ESO headquarter in 
Germany. 

The normalization of the spectra to the continuum level consisted of several 
steps described in detail by 
Hubrig et al.\ (\cite{Hubrig2013}). 
The Stokes~$I$ and $V$ parameters were derived following the ratio method 
described by 
Donati et al.\ (\cite{Donati1997}), 
and null polarization spectra were calculated  by combining the 
subexposures in such a way that the intrinsic source polarization
cancels out, yielding a diagnostic null $N$ spectrum. These steps
ensure that no spurious signals are present in the obtained data 
(e.g.\ Ilyin \cite{Ilyin2012}).


\section{Spectral variability}\label{sect:variab}

Not much is known about the presence of chemical spots on the surface of 
magnetic Herbig Ae stars, although they are usually considered as progenitors 
of classical magnetic Ap stars, which generally display a surface inhomogeneous 
distribution of different elements. In Ap stars with large-scale organized, 
predominantly dipolar magnetic fields, the non-uniformly distributed chemical 
elements show a certain symmetry with respect to the magnetic field 
configuration. As an example, rare earth element spots are frequently 
identified close to magnetic poles, while iron peak element spots tend to 
concentrate in the vicinity of the magnetic equator 
(e.g.\ Elkin et al.\ \cite{elkin2010}). 
The previous studies of HD\,101412 by 
Hubrig et al.\ (\cite{Hubrig2010}, \cite{Hubrig2011a}) 
using a sample of iron lines revealed line variations of equivalent widths, 
radial velocities, line widths, line asymmetries, and mean magnetic field 
modulus. Given the rather low SNR of the HARPS spectra, to study the spectral 
variability, we employed the  Least-Squares Deconvolution (LSD) technique 
allowing us to achieve a much higher SNR in the LSD spectra. The details of 
this technique can be found in the work of 
Donati et al.\ (\cite{Donati1997}). 

The absence of $\delta$~Scuti-like pulsations was already discussed in the 
previous work of J\"arvinen et al.\ (\cite{jarvinen}) 
where the authors verified that no changes in the line profile shape or radial velocity 
shifts are present in the obtained spectra on a short-time scale. This step 
was of importance since pulsations are known to have an impact on the analysis 
of the presence of a magnetic field and its strength 
(e.g.\ Schnerr et al.\ \cite{schnerr2006}; Hubrig et al.\ \cite{Hubrig2011b}). 

\begin{figure*}
\resizebox{\hsize}{!}{\includegraphics{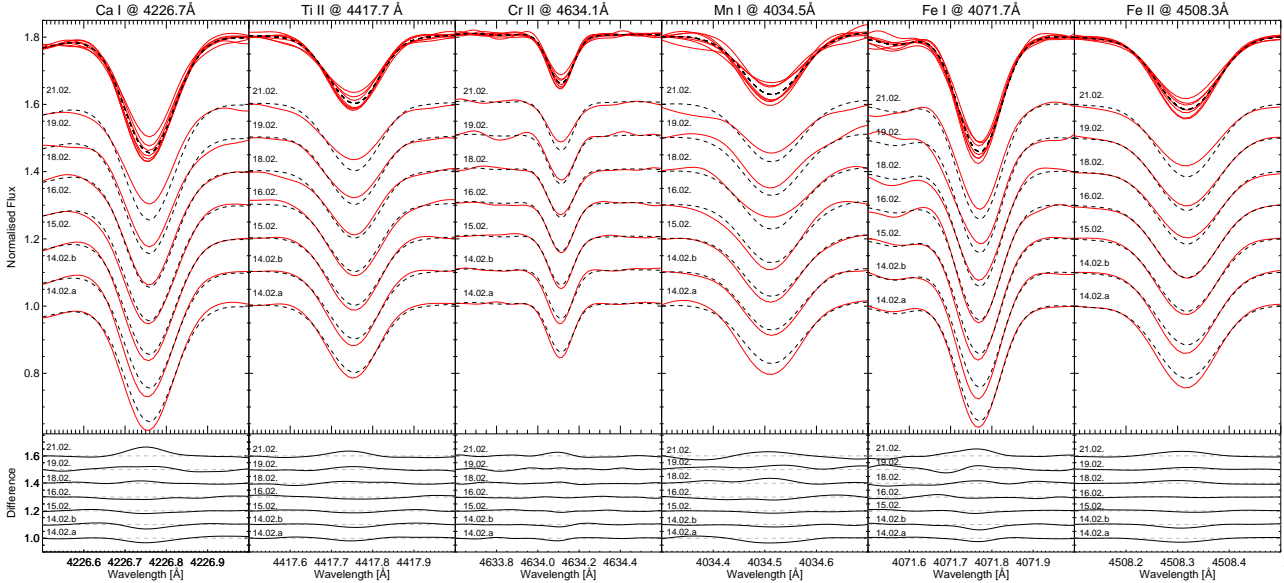}}
\caption{Comparison of Stokes~$I$ profiles for
spectral lines belonging to the elements Ca, Ti, Cr, Mn, and Fe recorded with 
HARPS on seven different epochs. The black dashed lines present the average 
profiles. The overplotted profiles are shown on the top, whereas the lower 
panels display the differences between the Stokes~$I$ profiles obtained for 
the individual exposures and the average Stokes~$I$ profile.}
\label{fig:nightly}
\end{figure*}

\begin{figure}
\centering
\includegraphics[width=0.32\textwidth]{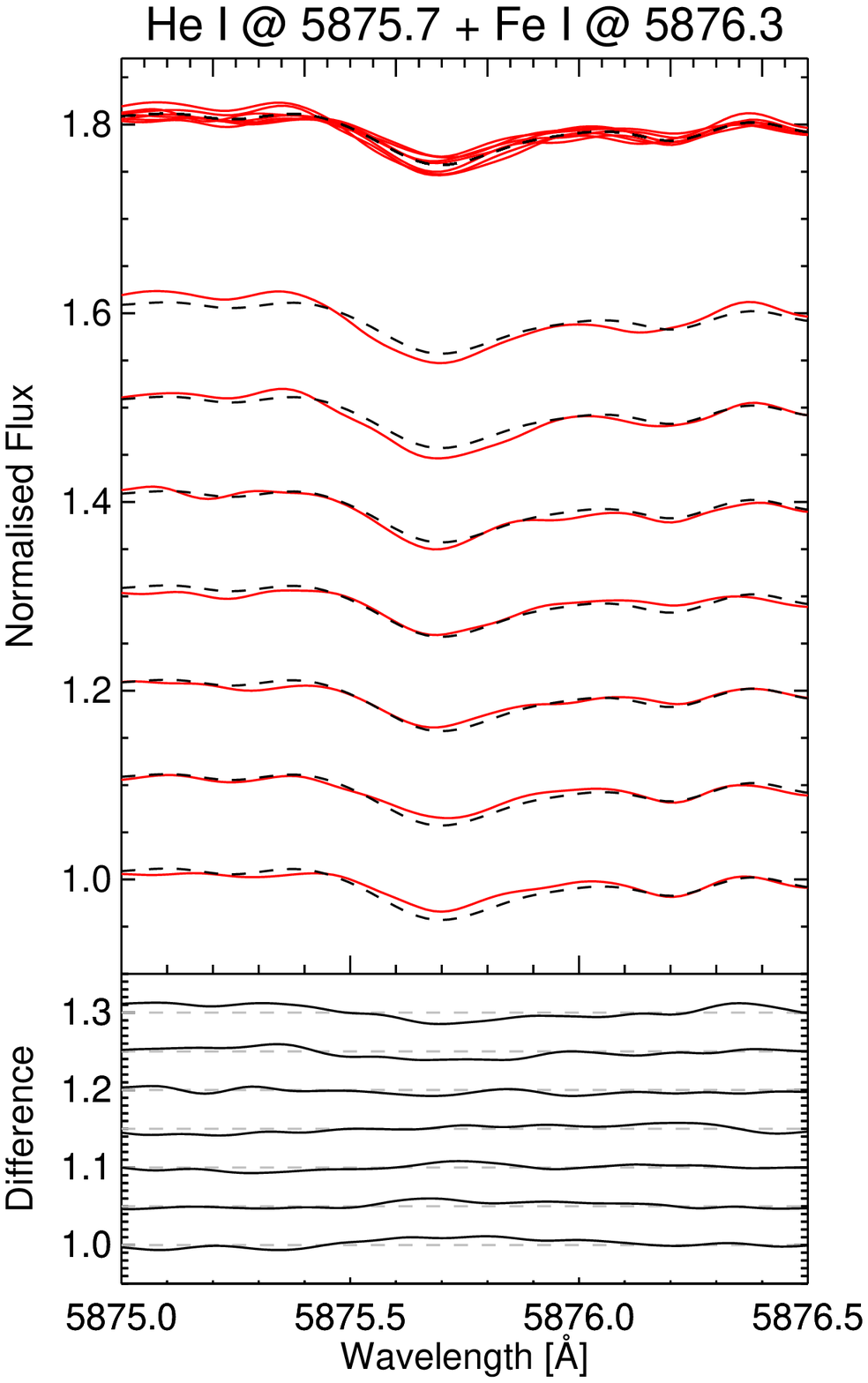}
\caption{Comparison of the Stokes~$I$ profiles for the \ion{He}{i}
$\lambda$ 5875.7 line recorded with HARPS on seven different epochs. As in 
Fig.~\ref{fig:nightly}, the black dashed lines present the average profiles. 
The overplotted profiles are shown on the top, whereas the lower panel 
displays the differences between the Stokes~$I$ profiles obtained for the 
individual exposures and the average Stokes~$I$ profile.}
\label{fig:nightly2}
\end{figure}

\begin{figure*}
\resizebox{\hsize}{!}{\includegraphics{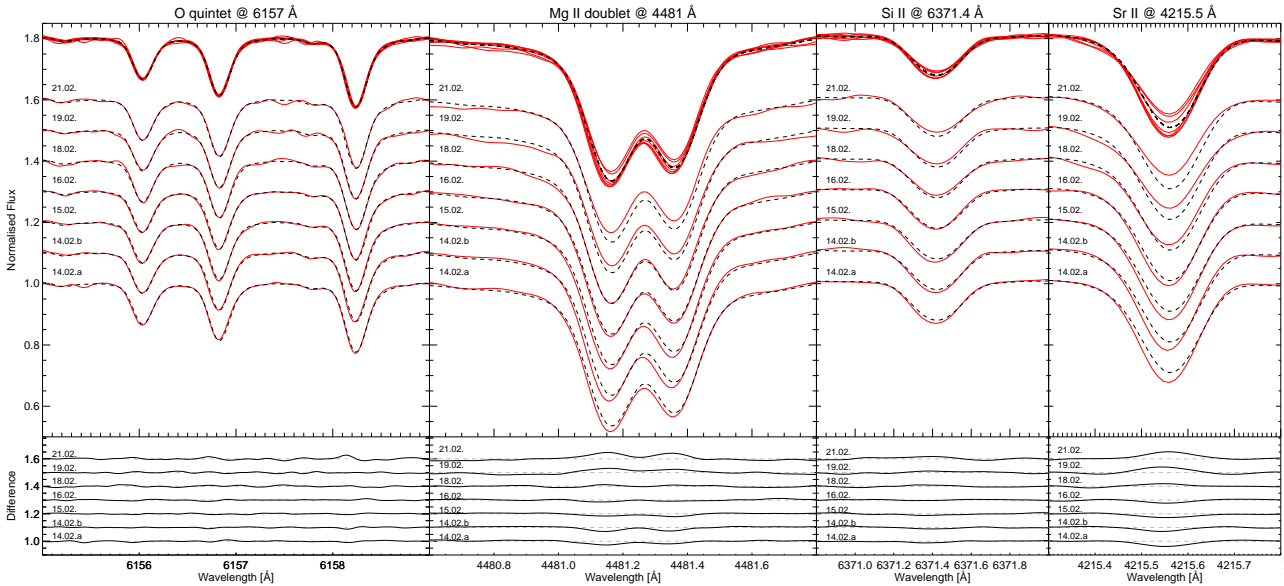}}
\caption{Comparison of Stokes~$I$ profiles for spectral lines belonging to 
the elements O, Mg, Si, and Sr recorded with HARPS on seven different epochs. 
The black dashed lines present the average profiles. As in 
Fig.~\ref{fig:nightly}, the overplotted profiles are shown on the top, whereas 
the lower panels display the differences between the Stokes~$I$ profiles 
obtained for the individual exposures and the average Stokes~$I$ profile.}
\label{fig:nightly1}
\end{figure*}

In the previous studies by 
Hubrig et al.\ (\cite{Hubrig2011a}, \cite{Hubrig2012}), 
the available spectra had an imperfect phase coverage over the
rotation cycle, especially in the phase range between 0.2 and 0.5 
(see Fig. 5 in Hubrig et al.\ \cite{Hubrig2011a}). 
The polarimetric HARPS spectra allow us now to partly close this gap
covering the phases from 0.131 to 0.297. For the study of the
variability of the line profiles belonging to different elements over
different rotation phases, we tried to choose only lines with a low
magnetic sensitivity, i.e.\ lines with low Land\'e factors. The reason
for this is that the magnetic field in HD\,101412 is rather strong, 
and magnetic intensification, i.e.\ differential broadening of spectral 
lines having different magnetic sensitivities is expected 
(e.g.\ Hubrig et al.\ \cite{Hubrig1999}). 
In Fig.~\ref{fig:nightly}, for different elements, we present a comparison of 
Stokes~$I$ profiles for spectral lines with low Land\'e factors obtained on 
seven different epochs. The studied spectral lines are 
\ion{Ca}{i} ($\lambda$ 4226.7\,\AA, $g_{\rm eff}=1.000$), 
\ion{Ti}{ii} ($\lambda$ 4417.7\,\AA, $g_{\rm eff}=0.795$), 
\ion{Cr}{ii} ($\lambda$ 4634.1\,\AA, $g_{\rm eff}=0.508$), 
\ion{Mn}{i} ($\lambda$ 4034.5\,\AA, $g_{\rm eff}=0.827$), 
\ion{Fe}{i} ($\lambda$ 4071.7\,\AA, $g_{\rm eff}=0.679$), and 
\ion{Fe}{ii} ($\lambda$ 4508.3\,\AA, $g_{\rm eff}=0.500$). 
Due to the presence of a rather strong noise in the HARPS spectra, before 
plotting the line profiles, we have slightly smoothed them using a Gaussian 
filter with a width of 0.01\,\AA{}. A clear decrease of line intensities is 
detected in all studied profiles in the time interval between the first and 
the last observing night. The variability displayed by the iron lines shows a 
trend similar to the iron line variability already discussed in the previous 
studies by 
Hubrig et al.\ (\cite{Hubrig2010}, \cite{Hubrig2011a}).
Since we also detect variations of the lines belonging to the elements 
Ca, Ti, Cr, and Mn as a function of the rotation phase, we conclude
that all these elements are inhomogeneously distributed on the surface 
of HD\,101412. The previously reported variability of a few other 
elements, such as He, Si, Mg, and Sr, by 
Hubrig et al.\ (\cite{Hubrig2011a,Hubrig2012}) 
is also detected in our HARPS spectra. In Figs.~\ref{fig:nightly2} and 
\ref{fig:nightly1}, we present the behaviour of the profiles of 
\ion{He}{i} $\lambda$ 5876, 
the oxygen quintet \ion{O}{i} at $\lambda$ 6157, 
\ion{Si}{ii} $\lambda$ 6371.4,  
the \ion{Mg}{ii} doublet at $\lambda$ 4481, 
and \ion{Sr}{ii} at $\lambda$ 4215.5. 
For these elements, due to the sparseness of the available lines in
the spectrum of HD\,101412, it was not possible to choose lines with
low Land\'e factors. Among them, we detect that He exhibits
variability that is opposite to the behaviour of the other studied
elements: the line intensity is increasing between the first and the
last observing nights. Also oxygen behaves differently, showing no 
changes of the line profiles over the same time interval.


\section{Magnetic field measurements}\label{sect:mf_meas}

Generally, in classical magnetic Ap stars with inhomogeneous surface element 
distribution, the actual measured magnetic field strength is dependent on the surface 
distribution of the elements from whose lines the measurements are made 
(e.g.\ Elkin et al.\ \cite{elkin2010}). 
Since the lines of different elements with different abundance distributions 
across the stellar surface sample the magnetic field in a different manner, 
we selected for our measurements four individual line masks containing lines 
belonging to the elements with the most rich spectra in HD\,101412. In our analysis, 
we have used the LSD method to boost up the signal in our HARPS spectra. This 
technique combines line profiles (using the assumption that line formation is similar 
in all lines) centred 
at the position of the individual lines given in the line mask and scaled 
according to the line strength and sensitivity to a magnetic field (i.e.\ 
to a Land\'e factor). The resulting average profiles (Stokes $I$, Stokes $V$,
and $N$) obtained by combining several lines, yield an increase 
in signal-to-noise ratio and therefore in sensitivity to polarization signatures. 
Only lines stronger than 10\% of the continuum considering only natural 
broadening have been selected. Following the study based on \ion{Fe}{i} lines presented by 
J\"arvinen et al.\ (\cite{jarvinen})
the line masks for the LSD code were created for atmospheric 
parameters $T_{\rm eff}=8\,300$\,K as well as $T_{\rm eff}=10\,000$\,K and 
$\log\,g=3.8$ using the VALD database 
(e.g.\ Kupka et al.\ \cite{kupka2000}).
Since line blends affect the individual lines, i.e.\ occurring at different 
positions within the line profile and having different strength, only 
unblended lines were chosen. The first pair of masks was consisting of 86
and 50 \ion{Ti}{ii} lines for $T_{\rm eff}=8\,300$\,K and $T_{\rm eff}=10\,000$\,K, 
respectively, and the second pair of 31 and 21 \ion{Cr}{ii} lines, again for 
the two temperatures. The results based on 52 \ion{Fe}{ii} lines forming at
$T_{\rm eff}=8\,300$\,K were already published in 
J\"arvinen et al.\ (\cite{jarvinen}),
and therefore only \ion{Fe}{ii} lines (47 in all) forming at $T_{\rm eff}=10\,000$\,K 
are considered here. The forth line mask included the lines of all these 
elements together forming at $T_{\rm eff}=10\,000$\,K. No line mask for He, Ca, Mn, 
or any rare earth element was created because of the low number of lines 
suitable for magnetic field measurements. Due to the especially low 
SNR at the blue and red ends of the HARPS spectra, the spectral 
lines used for the computation of the LSD profiles have been selected in the 
wavelength region from 4000 to 6500\,\AA. The mean longitudinal magnetic field 
is evaluated by computing the first-order moment of the Stokes~$V$ profile 
according to 
Mathys (\cite{Mathys1989}).

\begin{figure*}
\resizebox{\hsize}{!}{\includegraphics{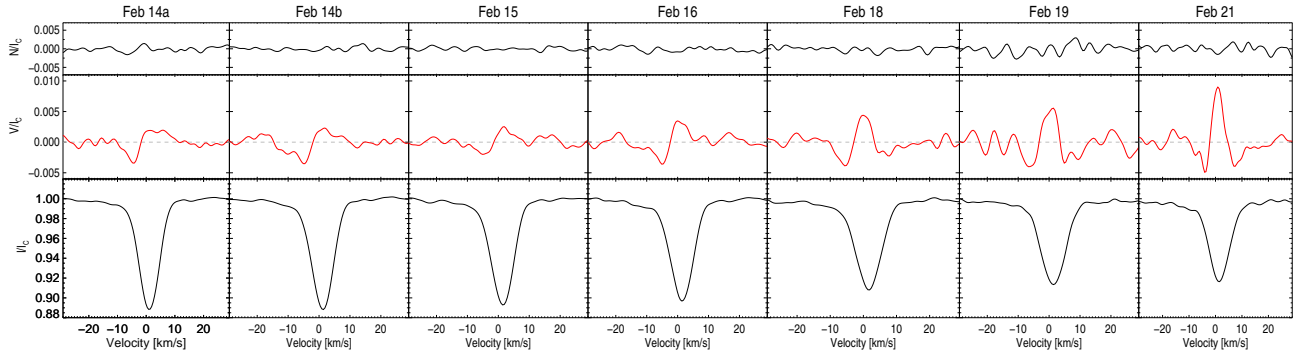}}
\caption{The LSD $I$, $V$, and $N$ spectra obtained for HD\,101412 on seven 
different epochs using a sample of 86 \ion{Ti}{ii} lines forming at $T_{\rm eff}=8\,300$\,K.}
\label{fig:LSDTiII}
\end{figure*}

\begin{figure*}
\resizebox{\hsize}{!}{\includegraphics{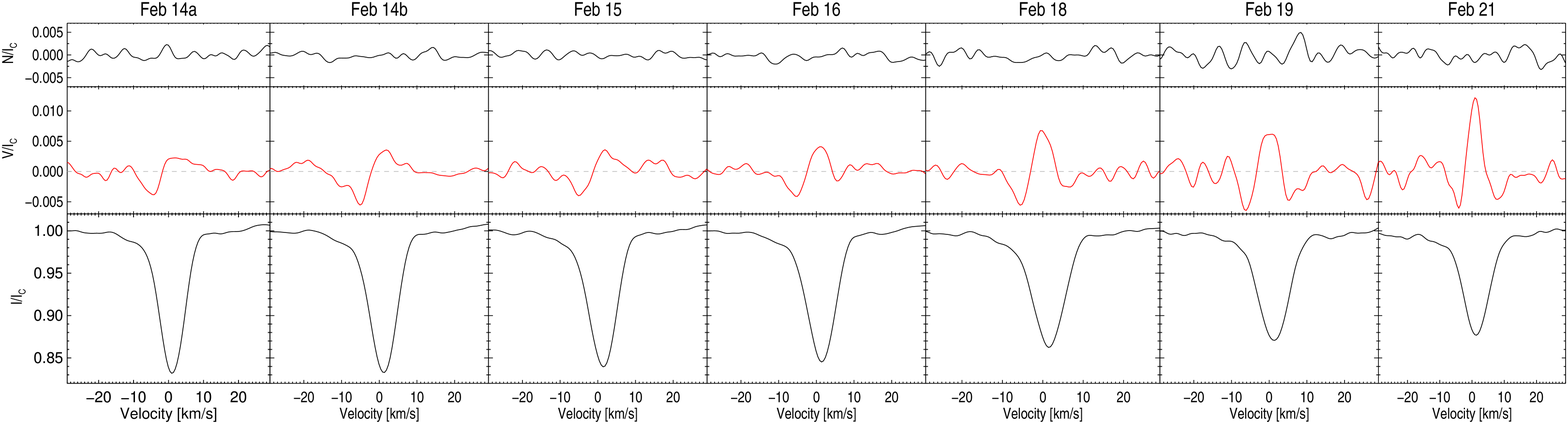}}
\caption{As Fig.~\ref{fig:LSDTiII} but using a sample of 50 \ion{Ti}{ii} lines 
forming at $T_{\rm eff}=10\,000$\,K.}
\label{fig:LSDTiII2}
\end{figure*}

\begin{figure*}
\resizebox{\hsize}{!}{\includegraphics{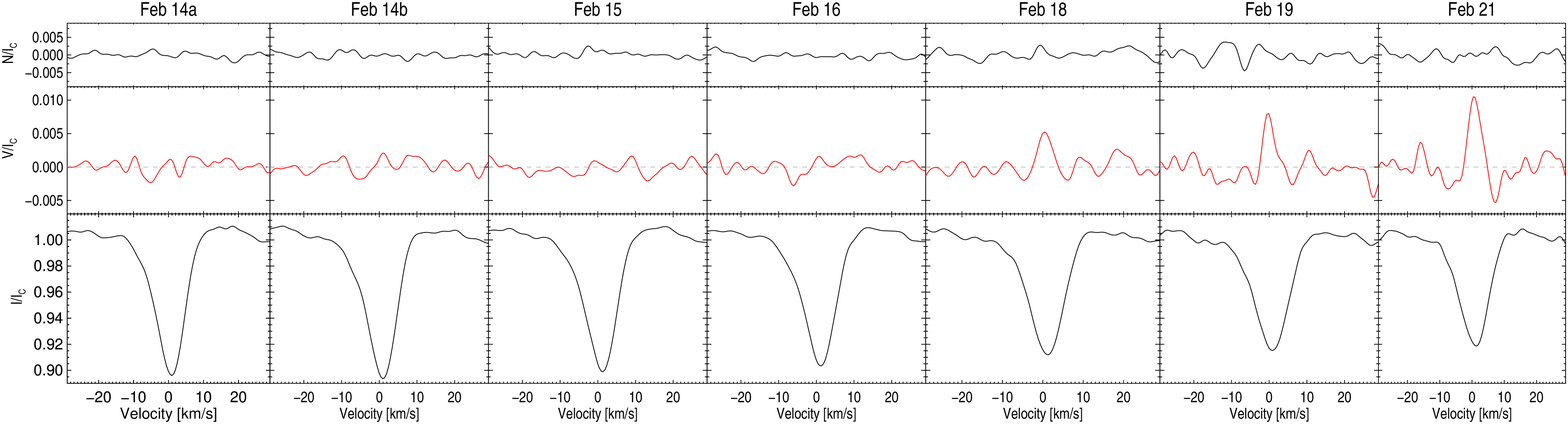}}
\caption{As Fig.~\ref{fig:LSDTiII}, but using a sample of 31 \ion{Cr}{ii} 
lines forming at $T_{\rm eff}=8\,300$\,K.}
\label{fig:LSDCrII}
\end{figure*}

\begin{figure*}
\resizebox{\hsize}{!}{\includegraphics{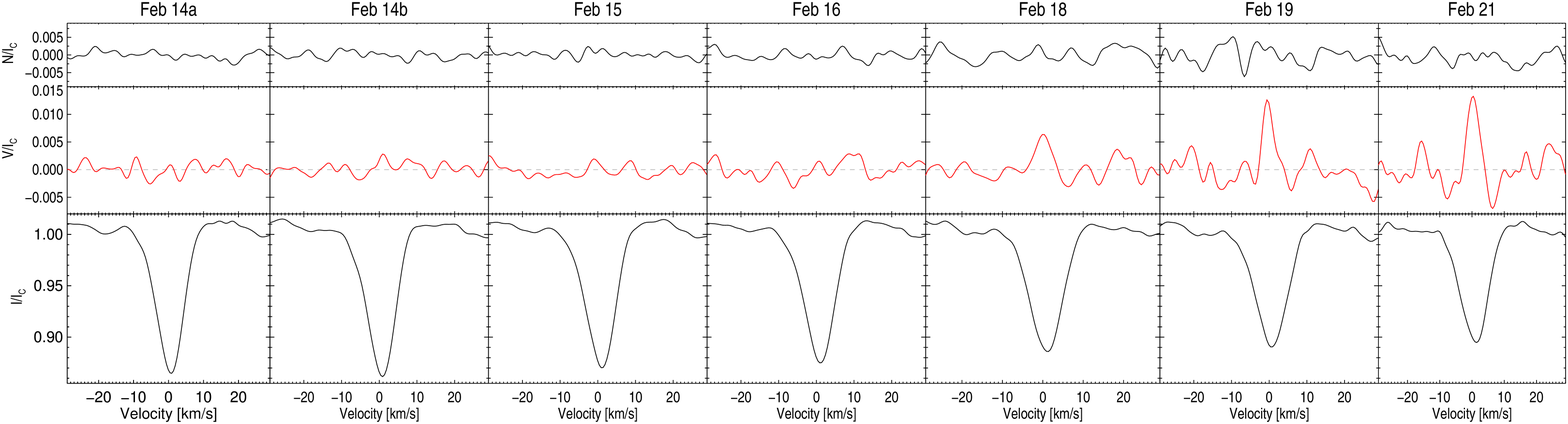}}
\caption{As Fig.~\ref{fig:LSDTiII}, but using a sample of 21 \ion{Cr}{ii} 
lines forming at $T_{\rm eff}=8\,300$\,K.}
\label{fig:LSDCrII2}
\end{figure*}

\begin{figure*}
\resizebox{\hsize}{!}{\includegraphics{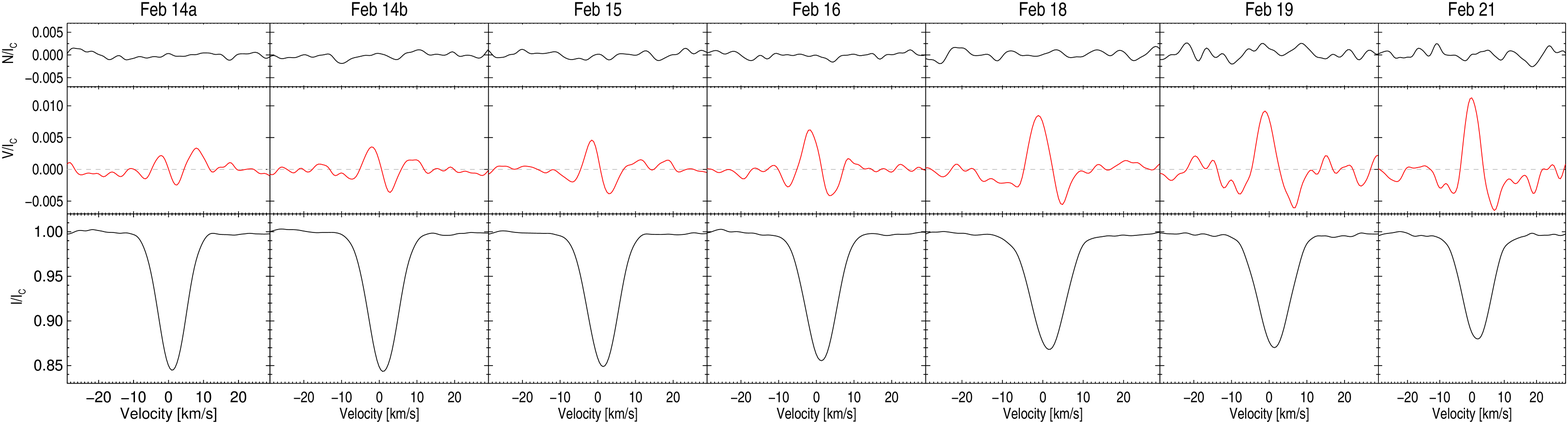}}
\caption{As Fig.~\ref{fig:LSDTiII}, but using a sample of 47 \ion{Fe}{ii} 
lines forming at $T_{\rm eff}=10\,000$\,K.}
\label{fig:LSDFeII2}
\end{figure*}

\begin{figure*}
\resizebox{\hsize}{!}{\includegraphics{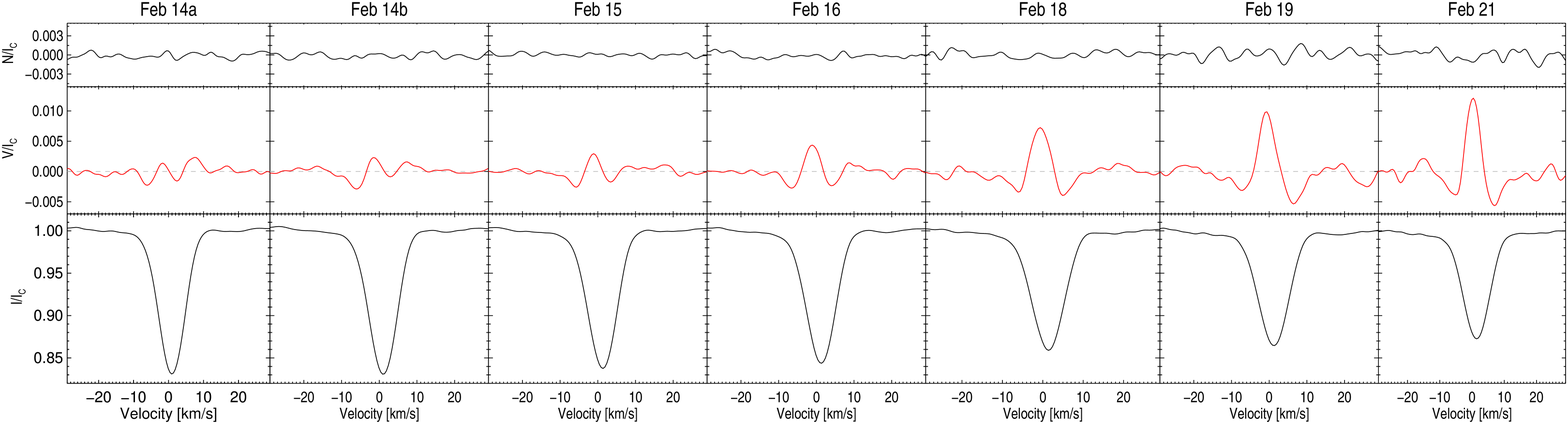}}
\caption{As Fig.~\ref{fig:LSDTiII}, but using all lines forming at $T_{\rm eff}=10\,000$\,K.}
\label{fig:LSDAll}
\end{figure*}

The resulting mean LSD Stokes~$I$, Stokes~$V$, and diagnostic $N$ profiles 
obtained for different line samples and different effective temperatures are presented in 
Figs.~\ref{fig:LSDTiII}--\ref{fig:LSDAll}. Distinct Zeeman features are 
detected at all epochs in all LSD Stokes~$V$ spectra. 
Only for the sample of \ion{Cr}{ii} lines on the first three epochs do the Stokes $V$ 
spectra not show any conspicuous feature, due to
the fact that the noise at these epochs has a stronger impact on the measurements caused by a
rather low number of lines in the line mask.

\begin{table*}
\centering
\caption{For each sample created for $T_{\rm eff}=10\,000$\,K , 
   the measured longitudinal magnetic field and
  the signal-to-noise value achieved in the LSD spectra are
  presented. The average Land\'e factor used in the LSD measurements
  is also indicated. The FAP is always less than $10^{-8}$.}
\label{tab:log_all}
\begin{tabular}{ccrr@{$\pm$}lrr@{$\pm$}lrr@{$\pm$}lrr@{$\pm$}l}
\hline\hline
HJD & Phase &
SNR & \multicolumn{2}{c}{$\left<B_{\rm z}\right>_{\rm Ti\,{II}}$}  &
SNR & \multicolumn{2}{c}{$\left<B_{\rm z}\right>_{\rm Cr\,{II}}$}  &
SNR & \multicolumn{2}{c}{$\left<B_{\rm z}\right>_{\rm Fe\,{II}}$}  &
SNR & \multicolumn{2}{c}{$\left<B_{\rm z}\right>_{\rm All}$}      \\
2450000+ & &
 & \multicolumn{2}{c}{[G]} &
 & \multicolumn{2}{c}{[G]} &
 & \multicolumn{2}{c}{[G]} &
 & \multicolumn{2}{c}{[G]} \\
\hline
 & &
\multicolumn{3}{c}{$\bar{g}_{\rm eff} = 1.04$} &
\multicolumn{3}{c}{$\bar{g}_{\rm eff} = 1.15$} &
\multicolumn{3}{c}{$\bar{g}_{\rm eff} = 1.09$} &
\multicolumn{3}{c}{$\bar{g}_{\rm eff} = 1.13$} \\
\hline
6337.801 & 0.131 & 670 & $-$99  & 18  & 302 & \multicolumn{2}{c}{} & 859  & $-$8  & 11  & 1215  & $-$69 & 8  \\ 
6337.835 & 0.132 & 749 & $-$122 & 18  & 350 & \multicolumn{2}{c}{} & 1203 & 5     & 14  & 1547  & $-$82 & 9  \\
6338.832 & 0.156 & 866 & $-$156 & 24  & 375 & \multicolumn{2}{c}{} & 1333 & 58    & 16  & 1835  & $-$72 & 12 \\
6339.709 & 0.176 & 458 & $-$115 & 20  & 295 & $-$87 & 35           & 725  & 62    & 18  & 1053  & $-$49 & 10 \\
6341.732 & 0.225 & 377 & $-$18  & 29  & 280 & 29    & 54           & 701  & 81    & 22  & 773   & $-$14 & 17 \\
6342.634 & 0.246 & 287 & 48     & 41  & 198 & $-$53 & 91           & 470  & 82    & 20  & 632   & 41    & 29 \\
6344.782 & 0.297 & 355 & 51     & 45  & 218 & $-$20 & 83           & 688  & 70    & 19  & 639   & 85    & 24 \\
\hline
\end{tabular}
\end{table*}

For all observations, the null spectra appear flat, indicating the absence of 
spurious polarization. Using the false alarm probability (FAP; 
Donati et al.\ \cite{Donati1992}) 
in the region corresponding to the whole Stokes~$I$ line profile (velocity range 
$\pm15$\,km\,s$^{-1}$), we obtain definite magnetic field detections with 
FAP~$<10^{-8}$ at all epochs for the samples of \ion{Ti}{ii}, \ion{Fe}{ii}, 
and the sample consisting of all lines together. For the sample 
of \ion{Cr}{ii} lines, definite detections are achieved only for the four last 
epochs. According to 
Donati et al.\ (\cite{Donati1992}), 
an FAP smaller than 10$^{-5}$ can be considered as a definite detection, while 
$10^{-5} < FAP < 10^{-3}$ is considered as a marginal detection. The 
measurement results are summarized in Table~\ref{tab:log_all}.

The inspection of the LSD profiles calculated for each sample presented in 
Figs.~\ref{fig:LSDTiII}--\ref{fig:LSDAll} reveals that the amplitude of the Zeeman 
features increases towards the latest epochs and that the majority of these 
features have a shape typical to that observed in classical magnetic stars 
during crossover, i.e.\ in the phase interval when the magnetic field changes 
its polarity 
(e.g.\ Hubrig et al.\ \cite{Hubrig2014}). 

J\"arvinen et al.\ (\cite{jarvinen}) already mentioned the fact that rather 
large differences in the shape of 
Zeeman features are observed between the \ion{Fe}{i} and \ion{Fe}{ii} samples 
on the first four epochs. In the LSD Stokes~$V$ profile obtained for 
\ion{Fe}{i} lines, they observed a distinct feature appearing as a 
second maximum in the red wing of the Zeeman profile. Furthermore, Stokes~$V$ 
profiles for neutral iron were slightly shifted to the blue by about 
4\,km\,s$^{-1}$. A similar behaviour of the LSD Stokes~$V$ profile was also 
observed for the sample containing the lines of different elements 
together and explained by the high proportion of neutral 
iron lines in this sample. As is shown in Figs.~\ref{fig:LSDTiII} and \ref{fig:LSDTiII2}, also 
the LSD Stokes~$V$ profiles 
obtained for the \ion{Ti}{ii} lines share the same behaviour. Nothing can be
concluded about the behaviour of the LSD Stokes~$V$ profiles for \ion{Cr}{ii} 
at the first four epochs due to the rather low SNR.

From our experience with the work on magnetic fields in classical Ap stars 
exhibiting chemical spots, we know that usually neutral and ionized Fe behave 
very similar, i.e.\ the magnetic field measurements for them are almost 
identical, indicating that these stars do not show temperature spots. Admittedly, 
the issue whether Herbig Ae stars exhibit temperature spots typical to those 
observed in T\,Tauri stars is currently unexplored. On the other hand, the 
velocity information, i.e.\ the presence of small shifts in the LSD Stokes $V$ 
profiles obtained for the neutral Fe and \ion{Ti}{ii} lists indicates that we probably observe a 
contamination by the surrounding warm circumstellar (CS) matter in the form of a wind. 
Such a scenario was already discussed by 
J\"arvinen et al.\ (\cite{jarvinen})
and would be in agreement with the previous finding of 
Hubrig et al.\ (\cite{Hubrig2009}) 
who reported a strong contamination of the UVES spectra of this star by weak lines 
of neutral and ionized iron (see their Fig. 6), where the lines of neutral 
iron are by far more numerous than those of ionized iron. 

To decrease the impact of the contamination by the CS contribution, 
J\"arvinen et al.\ (\cite{jarvinen})
studied the behaviour of the LSD Stokes~$V$ profiles calculated for a
sample of neutral iron lines forming at a significantly higher
effective temperature, i.e.\ assuming $T_{\rm eff}=10\,000$\,K, and
could show that at a significantly higher $T_{\rm eff}$ distinct
features in the red wings of the Zeeman profiles and velocity shifts 
completely disappear. However, the LSD Stokes~$V$ profiles calculated 
for the first epochs for the sample of \ion{Ti}{ii} lines formed at
different $T_{\rm eff}$ still resemble each other rather well, i.e.
the profiles remain almost unchanged. Such a behaviour found 
for \ion{Ti}{ii} lines can possibly be explained by the fact that the
sample contains a large number of \ion{Ti}{ii} lines with a low
excitation potential and thus the impact of the CS contamination still persists.

\begin{table}
\centering
\caption{Magnetic field measurements of the Herbig~Ae star HD\,101412 
using the moment technique. All quoted errors are 1$\sigma$ uncertainties.}
\label{tab:all}
\begin{tabular}{crr@{$\pm$}lr@{$\pm$}lr@{$\pm$}l}
\hline\hline
HJD & SNR & 
\multicolumn{2}{c}{$\left<B_{\rm z}\right>_{\rm Fe_{II}}$}    &  
\multicolumn{2}{c}{$\left<B_{\rm xover}\right>$}   & 
\multicolumn{2}{c}{$\left<B_{\rm q}\right>$} \\
{\small 2450000+} &  & 
\multicolumn{2}{c}{[G]} & 
\multicolumn{2}{c}{[G]} & 
\multicolumn{2}{c}{[G]} \\
\hline
6337.801 & 101 & $-$62 & 28 & 33  & 69  & 3087 & 204 \\
6337.835 & 103 & $-$37 & 27 & 4   & 73  & 2999 & 198 \\
6338.832 & 104 & $-$7  & 27 & 70  & 64  & 3153 & 205 \\
6339.709 &  92 & 36    & 29 & 115 & 83  & 3103 & 257 \\
6341.732 &  76 & 70    & 35 & 350 & 93  & 3232 & 287 \\
6342.634 &  58 & 160   & 56 & 425 & 131 & 3381 & 403 \\
6344.782 &  60 & 141   & 54 & 511 & 130 & 3579 & 430 \\
\hline
\end{tabular}
\end{table}

To obtain additional characteristics of the magnetic field in HD\,101412, we 
used the moment technique developed by 
Mathys (e.g.\ \cite{Mathys1991}). 
Importantly, this technique allows us not only the estimation of the mean 
longitudinal magnetic field, but also to prove the presence of crossover 
effect and of the quadratic magnetic field. It was already shown in the past 
that depending on the magnetic field geometry, even stars with rather weak 
longitudinal magnetic fields can exhibit strong crossover effects and kG 
quadratic fields 
(see e.g.\ Mathys \cite{Mathys1995}; Landstreet \& Mathys 
\cite{LandstreetMathys2000}; 
Mathys \& Hubrig \cite{MathysHubrig1997}, \cite{MathysHubrig2006}). 
The moment technique was applied to the same sample of unblended 45 
\ion{Fe}{ii} lines as selected for the LSD analysis. These measurements are 
presented in Table~\ref{tab:all}. For each line in this sample, the measured 
shifts between the line profiles in the left- and right-hand circularly 
polarized HARPS spectra are used in a linear regression analysis in the 
$\Delta\lambda$ versus $\lambda^2 g_{\rm eff}$ diagram, following the 
formalism discussed by 
Mathys (\cite{Mathys1991}, \cite{Mathys1994}). 
Similar to the LSD results, the longitudinal magnetic 
field shows negative polarity during the first epochs and positive polarity 
during the last epochs. The values for the field strength are in good 
agreement with those obtained using the LSD method within the measurement 
uncertainties. No significant magnetic fields could be determined from the 
null spectra. The measured quadratic magnetic field varies from about 3\,kG 
to 3.6\,kG over the rotation phase interval from 0.131 to 0.297. The crossover 
effect at a significance level larger than 3$\sigma$ is detected at the 
last three epochs in the rotation phase interval from 0.225 to 0.297.


\section{ Discussion}\label{sect:disc}

\begin{figure}
\centering
\includegraphics[width=0.45\textwidth]{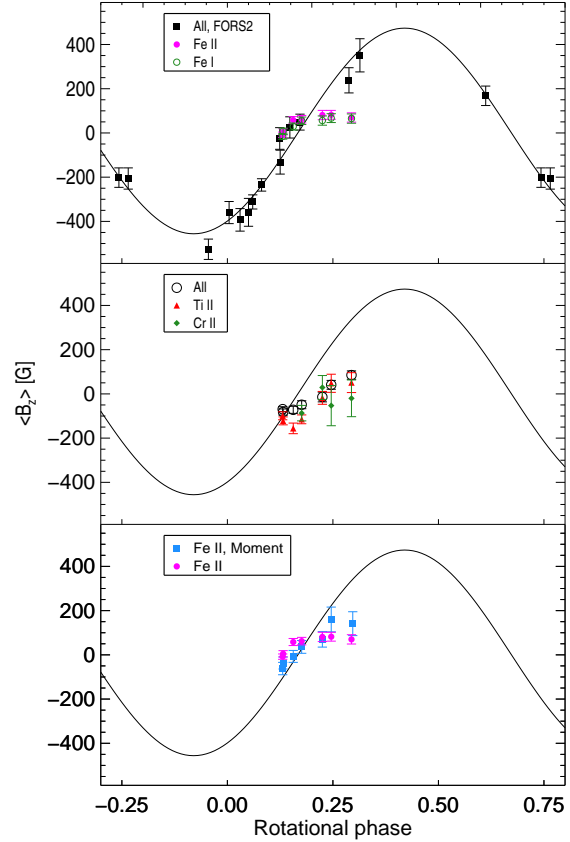}
\caption{Phase diagram for the longitudinal magnetic field measurements 
carried out using low-resolution FORS\,2 spectropolarimetric observations and 
high-resolution HARPS spectropolarimetric observations.
The fit on all three panels is from Hubrig et al.\ (\cite{Hubrig2011a}).
\emph{Top panel:} 
Black squares are from 
Hubrig et al.\ (\cite{Hubrig2011a}) 
based on FORS\,2 observations. 
Pink filled circles represent LSD measurements from \ion{Fe}{ii} lines that form 
at $T_{\rm eff}=10\,000$\,K, whereas green open circles are LSD measurements from 
\ion{Fe}{i} forming also at the same temperature (values are presented in 
J\"arvinen et al.\ \cite{jarvinen}).
\emph{Middle panel:} 
Open black circles show the LSD measurements obtained using all lines that form at 
$T_{\rm eff}=10\,000$\,K. Red filled triangles show \ion{Ti}{ii} line 
LSD measurements, and \ion{Cr}{ii} line LSD measurements are indicated with dark green 
filled diamonds.
\emph{Bottom panel:} 
Blue squares show measurements from \ion{Fe}{ii} lines using the moment 
technique and pink filled circles are, as in the top panel, \ion{Fe}{ii} 
LSD measurements.
}
\label{fig:Bz}
\end{figure}

In Fig.~\ref{fig:Bz}, we present the measurement results obtained for
different line masks using the LSD method and the results obtained
using the moment technique overplotted with the previous FORS\,2 
measurements. In the upper panel we present the measurements obtained by
Hubrig et al.\ (\cite{Hubrig2011a}) based on FORS\,2 observations together 
with our new measurements using \ion{Fe}{ii} lines formed 
at $T_{\rm eff}=10\,000$\,K and previous measurements using 
\ion{Fe}{i} forming at the same temperature, already presented in the work of 
J\"arvinen et al.\ \cite{jarvinen}). The distribution of the field values obtained 
using HARPSpol spectra is completely different compared to the magnetic field values
determined in  previous low-resolution FORS\,2 spectra, where hydrogen Balmer lines 
are the main contributors to the magnetic field measurements.
Similar differences between the distribution of field values obtained using HARPSpol and 
FORS\,2 are also detected in the measurements using samples of \ion{Ti}{ii} and 
\ion{Cr}{ii} lines presented in the middle panel.
Also the measurements using the moment technique presented in the bottom panel deviate 
significantly from the FORS\,2 measurements.  For all considered samples, the amplitude of the 
magnetic field strength is significantly lower than that observed in FORS\,2 spectra.

In contrast to 
low-resolution FORS\,2 measurements, the HARPS measurements are carried out 
on metal lines, which in classical magnetic Ap stars usually exhibit surface 
chemical spots. Our variability study of lines belonging to the elements 
Mg, Si, Ca, Ti, Cr, Mn, Fe, and Sr indicates the presence of changes in the line 
profiles over the part of the rotation cycle covered by HARPS observations, hinting at the 
presence of chemical spots on the stellar surface.
Furthermore, He seems to exhibit variability that is opposite to the behaviour of the other 
studied elements. The shape of the LSD Stokes $V$ profiles displaying crossover 
effect indicates that we observe a change from negative to 
positive polarity, i.e.\ both negative and positive magnetic poles are 
visible. The obtained result that the amplitude of the magnetic field strength 
measured on metal lines is 
significantly lower than that observed in FORS\,2 spectra is a well-known phenomenon 
frequently observed in Ap stars if the magnetic field measurements are 
carried out using line samples belonging to elements concentrated far from 
magnetic poles. For instance, in a number of studies of such stars, iron peak 
elements appear enhanced along the magnetic equator 
(e.g.\ Nesvacil et al.\ \cite{nesvacil2012}).
We note that the assumption of the presence of concentration of these elements 
in the region of the magnetic equator, similar to the iron peak element 
distribution appearing in Ap stars, would explain the amplitude and the shape 
of the observed Zeeman features displaying crossover effect. 
The results of our measurements of the mean longitudinal magnetic field based 
on different samples are comparable within the measurement accuracies using 
the LSD method.

J\"arvinen et al.\ (\cite{jarvinen}) revealed significant differences between 
the appearance of Zeeman features
using the sample of \ion{Fe}{ii} lines and \ion{Fe}{i} lines formed at 
$T_{\rm eff}=8\,300$\,K. This difference disappears if the LSD profiles are 
calculated for the sample of \ion{Fe}{i} lines at $T_{\rm eff}=10\,000$\,K. 
J\"arvinen et al.\ discuss that the  differences between the measurements using iron
line samples can be caused by temperature inhomogeneities or by a contamination by 
the surrounding warm CS matter. If temperature 
spots are indeed present on the surface of Herbig Ae stars, then it could be 
suggested that \ion{Fe}{i} lines are formed in somewhat cooler spots. On the 
other hand, the presence of small shifts of Stokes $V$ profiles to shorter 
wavelengths by about 4\,km\,s$^{-1}$ obtained for neutral Fe indicates that we 
probably observe a contamination by the surrounding warm CS matter in the form 
of a wind. The observed similarity between the 
shapes of Zeeman features in the Stokes $V$ spectra of \ion{Fe}{i} at 
$T_{\rm eff}=10\,000$\,K and \ion{Fe}{ii} support the suggested scenario.  
In this work we show for the first time that similar distinct features and shifts appear also 
in Zeeman profiles calculated for the sample of \ion{Ti}{ii} lines.

Magnetic fields in a few Herbig Ae stars were already discovered a decade ago 
(e.g.\ Hubrig et al.\ \cite{Hubrig2004}, \cite{Hubrig2006}, \cite{Hubrig2015}; 
Wade et al.\ \cite{Wade2005}), 
but our understanding of the interaction between the central stars, their 
magnetic fields, and their protoplanetary discs is still very limited. It is 
not clear yet whether the majority of Herbig Ae stars do possess weak magnetic 
fields, which are not detected due to the rather low accuracy of the magnetic 
field measurements in previous studies. To gain better knowledge of typical 
magnetic field strengths in Herbig Ae/Be stars, Hubrig et al.\ 
(\cite{Hubrig2015}) recently compiled all 
magnetic field measurements reported in previous spectropolarimetric studies. 
The obtained density distribution of the rms longitudinal magnetic field 
values reveals that only very few stars have fields stronger than 200\,G, and 
half of the sample possesses magnetic fields of about 100\,G and less. 
Noteworthy, in the currently largest high-resolution spectropolarimetric 
survey of the magnetic field in these stars by 
Alecian et al.\ (\cite{alecian2013})
with 132 measurements for 70 Herbig stars, the measurement accuracy is worse 
than 200\,G for 35\% of the measurements, and for 32\% of the measurements it 
is between 100 and 200\,G, i.e.\ only 33\% of the measurements showed a 
measurement accuracy of better than 100\,G.

We note also that without acquiring high-quality spectropolarimetric material, 
it will not be possible to decide whether there is a fundamental 
difference between magnetic and non-magnetic Herbig stars. Since the number of 
Herbig stars with a firmly detected magnetic field is currently small, 
fundamental questions related to the accretion geometry remain unanswered. The 
detection of very weak longitudinal magnetic fields in these stars put into 
question our current understanding of the magnetospheric accretion process in 
intermediate-mass pre-main sequence stars, as they indicate that their magnetic 
fields are by far weaker than those measured in their lower mass counterparts, 
the classical T\,Tauri stars, usually possessing kG magnetic fields. In this respect, the presence of 
a strong magnetic field in HD\,101412 and its slow rotation make this object a 
most favourable candidate for studies of the role of the magnetic field in the 
star formation process.


\acknowledgements
We would like to thank G.~Lo~Curto for his support with the reduction of the 
HARPS polarimetric spectra. 
SPJ acknowledges the support by the Deutsche Forschungsgemeinschaft, 
grant JA 2499/1~-1.


\end{document}